\documentclass[aps,twocolumn]{revtex4-2}
\usepackage[dvipdfmx]{graphicx}
\usepackage[dvipdfmx]{hyperref}

\newcommand{\be}{\begin{eqnarray}}
\newcommand{\ee}{\end{eqnarray}}
\newcommand{\no}{\nonumber}
\newcommand{\ba}{\begin{array}}
\newcommand{\ea}{\end{array}}
\newcommand{\bmat}{\left(\begin{array}}
\newcommand{\emat}{\end{array}\right)}

\newcommand{\e}{\mathrm e}

\begin{document}

\title{Generalized speed limits for classical stochastic systems 
and their applications to relaxation, annealing, and pumping processes}
\author{Kazutaka Takahashi}
\author{Yasuhiro Utsumi}
\affiliation{Department of Physics Engineering, Faculty of Engineering,
Mie University, Mie 514--8507, Japan}
\date{\today}

\begin{abstract}
We extend the speed limit of a distance between two states
evolving by different generators for quantum systems
[K. Suzuki and K. Takahashi, 
\href{https://doi.org/10.1103/physrevresearch.2.032016}
{Phys. Rev. Res. {\bf 2}, 032016(R) (2020)}]
to the classical stochastic processes described by the master equation.
We demonstrate that the trace distance between arbitrary evolving states 
is bounded from above by using a geometrical metric.
The geometrical bound reduces to the Fisher information metric
for the distance between the time-evolved state and the initial state.
We compare the bound in relaxation and annealing processes  
with a different type of bound
known for nonequilibrium thermodynamical systems.
For dynamical processes such as annealing and pumping processes, 
the distance between the time-evolved state and
the instantaneous stationary state becomes a proper choice and
the bound is represented by the Fisher information metric
of the stationary state.
The metric is related to the counterdiabatic term 
defined from the time dependence of the stationary state.
\end{abstract}
   
\maketitle

\section{Introduction}

It is widely known that 
a certain type of inequalities on time evolution 
generally holds in dynamical systems.
When the state of the system is described by a time-evolution law,
a distance measure for the time-evolved state and the initial state 
is bounded from above by using a geometrical metric.
This speed-limit inequality was first studied  
for closed quantum systems and is known as 
the Mandelstam--Tamm relation~\cite{MT}.
From a general perspective, it was recently understood that
the speed-limit inequality holds not only for quantum systems 
but also for classical stochastic systems~\cite{SCMdC, OO}.
The inequality was further extended to
nonequilibrium thermodynamic processes~\cite{SFS}.
The state distance is bounded from above by  
nonequilibrium quantities such as the entropy production.
The thermodynamic speed limit is one of recent interests on
nonequilibrium thermodynamics and has been discussed intensively.
Thus, due to increasing interests on general properties of dynamical processes 
including quantum controls, thermodynamical operations, 
computational cost, and information processing,
the speed limit has attracted renewed attention 
in recent years~\cite{DC}.

Suppose that the state of a dynamical system is characterized by $\rho(t)$.
It evolves as a function of time $t$ according to the time-evolution law.
When the difference between the time-evolved state and the initial state 
is characterized by a distance measure $D(\rho(t),\rho(0))$, 
we generally have an inequality 
\be
 D(\rho(t),\rho(0))\le \int_0^t ds\, v(s), \label{sl0}
\ee
and the maximum velocity $v(s)$ is determined from 
the geometrical property of the time evolution.
For Schr\"odinger dynamics, 
when the distance measure is defined by the Fubini-Study metric, 
the velocity $v(s)$ is represented by the variance of the energy.
This relation was generalized to give a similar bound 
for a distance between arbitrary time-evolved states 
$D(\rho^{(1)}(t),\rho^{(2)}(t))$~\cite{ST, HT}.

The main aim of the present study is 
to understand related formulas
for classical stochastic systems systematically.
We assume the Markovian dynamics and 
the time evolution is described by the master equation.
As we mentioned above,
the speed-limit inequalities for classical systems were discussed in
previous studies~\cite{SCMdC, OO, SFS}.
Each study focused especially on 
the use of the Wigner function to describe 
the quantum-classical correspondence~\cite{SCMdC},
the formal derivations of inequalities 
for several kinds of classical stochastic equations~\cite{OO}, 
and the representation of the bound by 
thermodynamic quantities~\cite{SFS}.
In the present study, we reformulate the relation from 
a geometrical perspective and generalize it to 
a distance between arbitrary time-evolved states.

The classical state is characterized by the probability distribution
with nonnegative variables,
which is contrasted to the quantum state where 
each component is characterized by a complex variable.
In addition, the master equation basically describes 
a relaxation to the stationary state.
These properties are strongly reflected to the tightness of the bound.
We need to study the bound more closely in specific examples, 
which is one of our aims in the present study.

The organization of the paper is as follows.
First, in Sec.~\ref{sec:sl}, we derive general inequality relations.
Next, we apply the derived inequalities to 
relaxation and annealing processes 
in Secs.~\ref{sec:app1} and \ref{sec:app2}, 
and pumping processes in Sec.~\ref{sec:app3}.
Finally, we give a summary in Sec.~\ref{sec:summary}.

\section{Speed limits for classical stochastic processes}
\label{sec:sl}

\subsection{General results}

We discuss a speed-limit inequality generally written 
for a distance measure $D$ as 
\be
 D(\rho^{(1)}(t),\rho^{(2)}(t))\le \int_0^t ds\, v_{12}(s), \label{sl}
\ee
where $\rho^{(1)}(t)$ and $\rho^{(2)}(t)$ represent arbitrary states. 
The time evolution of each state $\rho^{(i)}(t)$, 
where $i$ takes one or two, is described by a continuous Markov process
\be
 \dot{\rho}^{(i)}(t)=K^{(i)}[\rho^{(i)}(t)],
\ee
where the dot symbol denotes the time derivative.
The explicit form of $K^{(i)}[\rho]$ depends on the detail of the system.
We also set the initial condition $\rho^{(1)}(0)=\rho^{(2)}(0)$.
Then, $v_{12}(t)$ in Eq.~(\ref{sl}) can be obtained by considering  
the infinitesimal time evolution.
We set $D(\rho^{(1)}(t),\rho^{(2)}(t))=0$ to write 
\be
 D(\rho^{(1)}(t+dt),\rho^{(2)}(t+dt))=v_{12}(t)dt+O(dt^2). \label{v12}
\ee
$v_{12}(t)$ is a nonnegative quantity and has a meaning of 
a state distinguishable metric.
Equation (\ref{sl}) is a generalization of Eq.~(\ref{sl0}) and 
can be useful when we want to know the behavior of
the unknown state $\rho^{(1)}(t)$ from the well-known state $\rho^{(2)}(t)$.
Then, it is important to represent $v_{12}(t)$ 
with respect to $\rho^{(2)}(t)$ only.

The above formulation is applied to any quantum and classical systems.
The applications to Schr\"odinger dynamics were 
discussed in Refs.~\cite{ST, HT, Takahashi22}.
In the present study, we treat classical stochastic processes.
The state of the system is described by the probability distribution.
We use the vector representation 
\be
 |p(t)\rangle 
 =\sum_{n=1}^N|n\rangle p_n(t)
 =\bmat{c} p_1(t) \\ p_2(t) \\ \vdots \\ p_N(t) \emat,
\ee
where $N$ denotes the number of events.
Each component $p_n(t)=\langle n|p(t)\rangle$ ($n=1,2,\dots,N$) 
is a nonnegative quantity and the normalization condition 
$\sum_{n=1}^Np_n(t) =1$ is imposed.
We compare two different states $|p^{(1)}(t)\rangle$ and $|p^{(2)}(t)\rangle$
with the initial condition $|p^{(1)}(0)\rangle=|p^{(2)}(0)\rangle$.
The time evolution of each state is described by 
the classical master equation 
\be
 |\dot{p}^{(i)}(t)\rangle=W^{(i)}(t)|p^{(i)}(t)\rangle \label{pi}
\ee
with $i=1$ or 2.
The offdiagonal components of the transition-rate matrix $W^{(i)}(t)$,
$\langle m|W^{(i)}(t)|n\rangle$ with $m\ne n$, 
are nonnegative and the diagonal components,
$\langle n|W^{(i)}(t)|n\rangle$, are determined from the relation
$\sum_{n=1}^N\langle n|W^{(i)}(t)=0$.

The explicit form of $v_{12}(t)$ depends on the definition of 
the distance measure. 
When we use the trace distance 
\be
 D(p^{(1)}(t),p^{(2)}(t))=\frac{1}{2}\sum_{n=1}^N
 \left|p^{(1)}_n(t)-p^{(2)}_n(t)\right|,
\ee
we obtain by using Eq.~(\ref{v12}) 
\be
 & & D(p^{(1)}(t),p^{(2)}(t)) \le \frac{1}{2}\sum_{n=1}^N\int_0^t ds\, \no\\
 & & \times \left|\langle n|
 \left(W^{(1)}(s)-W^{(2)}(s)\right)|p^{(2)}(s)\rangle \right|. 
 \label{csl}
\ee
We can also use the Cauchy--Schwartz inequality 
$\sum_na_nb_n\le \sqrt{\sum_n a_n^2\sum_n b_n^2}$ to bound 
the right hand side from above as 
\be
 & & D(p^{(1)}(t),p^{(2)}(t)) \le
 \frac{1}{2}\int_0^t ds\,
 \Biggl[\sum_{n=1}^Np_n^{(2)}(s) \no\\
 && \times 
 \left(\frac{
 \langle n|\left(W^{(1)}(s)-W^{(2)}(s)\right)|p^{(2)}(s)\rangle}{p_n^{(2)}(s)}
 \right)^2\Biggr]^{1/2}.
 \label{cslf}
\ee
This bound can also be derived by using 
the Bhattacharyya angle as a distance measure.
It is defined as 
\be
 \Theta(p^{(1)}(t),p^{(2)}(t))=\arccos\sum_{n=1}^N 
 \sqrt{p_n^{(1)}(t)p^{(2)}_n(t)}.
\ee
By using the infinitesimal time evolution, we can find 
that $\Theta(p^{(1)}(t),p^{(2)}(t))$ is bounded from above by the quantity 
of the right hand side in Eq.~(\ref{cslf}).
Since we generally have the relation 
$D(p^{(1)}(t),p^{(2)}(t))\le \Theta(p^{(1)}(t),p^{(2)}(t))$,
we can obtain Eq.~(\ref{cslf}) as a result.

Equations (\ref{csl}) and (\ref{cslf}) are the main general results
of the present study.
We note that $|p^{(2)}(s)\rangle$ on the right hand side of these equations
can be replaced by $|p^{(1)}(s)\rangle$.
We can use the convenient choice depending on the problem.
Although Eq.~(\ref{cslf}) gives a loose bound 
compared to that in Eq.~(\ref{csl}), 
we find in the following discussion 
that the bound in Eq.~(\ref{cslf}) is used
to obtain some geometrical understanding.

To support the heuristic derivation of Eq.~(\ref{csl}), 
we discuss an explicit derivation in the following.
It is also used to find a different form of the bound.
We start from the integral representation
\be
 && |p^{(1)}(t)\rangle = |p^{(2)}(t)\rangle \no\\
 && +\int_0^t ds\,U^{(1)}(t,s)(W^{(1)}(s)-W^{(2)}(s))|p^{(2)}(s)\rangle, \label{int}
\ee
where the time evolution operator $U^{(1)}(t,s)$ 
is defined from the relation $|p^{(1)}(t)\rangle=U^{(1)}(t,s)|p^{(1)}(s)\rangle$ 
and satisfies
$\partial_t U^{(1)}(t,s)=W^{(1)}(t)U^{(1)}(t,s)$,
$\partial_s U^{(1)}(t,s)=-U^{(1)}(t,s)W^{(1)}(s)$, 
and $U^{(1)}(s,s)=1$.
By taking the time derivative of Eq.~(\ref{int}), we confirm 
that the equation is consistent with Eq.~(\ref{pi}).
It is also consistent with the initial condition
$|p^{(1)}(0)\rangle=|p^{(2)}(0)\rangle$.
These properties justify Eq.~(\ref{int}).
We can write the trace distance as
\be
 & & D(p^{(1)}(t),p^{(2)}(t)) = \frac{1}{2}\sum_{m=1}^N \left|
 \int_0^t ds\,\langle m|U^{(1)}(t,s)\right. \no\\
 && \times 
 (W^{(1)}(s)-W^{(2)}(s))|p^{(2)}(s)\rangle
 \biggr|. 
 \label{du}
\ee
We insert $\sum_{n=1}^N|n\rangle\langle n|=1$ 
between the first line and the second line of this equation 
and use the general inequality 
$\left|\sum_n\int_0^tds\,a_n(s)\right|\le \sum_n\int_0^tds\left|a_n(s)\right|$
to obtain
\be
 D(p^{(1)}(t),p^{(2)}(t)) &\le& \frac{1}{2}\sum_{m,n=1}^N \int_0^t ds\,
 \left|\langle m|U^{(1)}(t,s)|n\rangle\right. \no\\
 && \times\left.\langle n|(W^{(1)}(s)-W^{(2)}(s))|p^{(2)}(s)\rangle
 \right|. \no\\
 \label{u}
\ee
Since $\langle m|U^{(1)}(t,s)|n\rangle$ represents 
a probability, it is a nonnegative quantity 
and satisfies $\sum_{m=1}^N\langle m|U^{(1)}(t,s)|n\rangle=1$.
Thus, we obtain the bound in Eq.~(\ref{csl}).

Using the exact relation in Eq.~(\ref{du}), 
we can obtain a different bound~\cite{BFGY}.
We define 
\be
 K(t) = \int_0^t ds\,(W^{(1)}(s)-W^{(2)}(s)),
\ee
and use the integration by parts as 
\be
 && |p^{(1)}(t)\rangle=|p^{(2)}(t)\rangle
  +\int_0^t ds\,U^{(1)}(t,s)\frac{dK(s)}{ds}|p^{(2)}(s)\rangle \no\\
 && = |p^{(2)}(t)\rangle
 +K(t)|p^{(2)}(t)\rangle +\int_0^t ds\,U^{(1)}(t,s)\no\\
 &&\times\left(W^{(1)}(s)K(s)-K(s)W^{(2)}(s)\right)
 |p^{(2)}(s)\rangle. 
 \label{int2}
\ee
Then, by using the same procedure as Eq.~(\ref{u}), 
we can eliminate $U^{(1)}$ as 
\be
 D(p^{(1)}(t),p^{(2)}(t)) &\le& 
 \frac{1}{2}\sum_{n=1}^N\left|\langle n|K(t)|p^{(2)}(t)\rangle\right| \no\\
 && +\frac{1}{2}\sum_{n=1}^N\int_0^t ds\,\left|
 \langle n|\left(W^{(1)}(s)K(s) \right.\right.
 \no\\ & &
 \left.\left.-K(s)W^{(2)}(s)\right)|p^{(2)}(s)\rangle\right|. 
 \label{csl2}
\ee
This is a different representation of the bound of $D(p^{(1)}(t),p^{(2)}(t))$.
Compared to Eq.~(\ref{csl}), 
we expect that the right hand side of Eq.~(\ref{csl2}) becomes smaller 
when the transition-rate matrix is an oscillating function of time 
and satisfies $K(t)=0$.
Since the last term of the right hand side in Eq.~(\ref{int2}) 
takes a similar form as that in Eq.~(\ref{int}), 
we can repeat the same calculation any number of times 
by changing the definition of $K(t)$ 
to obtain different forms of the bound.

\subsection{Relaxation processes}

The derived inequalities, Eqs.~(\ref{csl}), (\ref{cslf}), and (\ref{csl2}),
are applied to arbitrary sets of 
probability distributions, $|p^{(1)}(t)\rangle$ and $|p^{(2)}(t)\rangle$.
The master equation basically describes a relaxation process.
In that case, the time-evolved state is compared to the initial state
as in the standard speed-limit analysis.
By putting $(p^{(1)}(t),p^{(2)}(t))=(p(0),p(t))$ 
and $(W^{(1)}(t),W^{(2)}(t))=(0,W(t))$ for the general inequality 
derived in the previous subsection, we obtain 
\be
 D(p(t),p(0)) &=& \frac{1}{2}\sum_{n=1}^N
 \left|\int_0^t ds\,\langle n|W(s)|p(s)\rangle\right| \no\\
 &\le& \frac{1}{2}\sum_{n=1}^N\int_0^t ds\,
 \left|\langle n|W(s)|p(s)\rangle\right| \no\\
 &\le& \frac{1}{2}\int_0^t ds\,\sqrt{J(s)}.
 \label{csl0}
\ee
The second line corresponds to the general relation in Eq.~(\ref{csl})
and the third line to Eq.~(\ref{cslf}).
$J(t)$ represents the Fisher information metric defined as 
\be
 J(t)=\sum_{n=1}^N p_n(t)\left(\frac{\dot{p}_n(t)}{p_n(t)}\right)^2.
\ee
The representation of the bound by the Fisher information metric
is well known as the form 
$\Theta(p(t),p(0))\le \frac{1}{2}\int_0^t ds\,\sqrt{J(s)}$.
The Fisher information metric is a geometrical quantity and is interpreted
as the squared velocity in the probability distribution space.
The bound by the Fisher information metric 
is known as the thermodynamic speed limit.
The statistical distances are defined in quantum and classical stochastic
systems and their bounds are characterized by 
a Riemannian metric~\cite{Wootters81, SNB85, BC94, Ruppeiner95}.
The thermodynamic speed limit has been discussed recently 
in the context of the information geometry~\cite{Ito18, ID20}.

It is instructive to see how the bound is dependent on $t$.
To treat relaxation processes, 
we consider the case that 
the transition-rate matrix $W$ is time independent.
We write the transition-rate matrix in the spectral representation 
\be
 W=\sum_{n=1}^{N}\Lambda_n|R_n\rangle\langle L_n|.\label{spectral}
\ee
The right eigenstate $|R_n\rangle$ and 
the left eigenstate $\langle L_n|$ satisfy 
$\langle L_m|R_n\rangle=\delta_{m,n}$ and 
$\sum_{n=1}^N|R_n\rangle\langle L_n|=1$.
The eigenvalue $\Lambda_n$ takes a nonpositive value.
One of them, we set it as $\Lambda_1$,  
must be equal to zero and the corresponding eigenstate $|R_1\rangle$
represents the stationary state.
We assume the stationary state is unique, which implies 
$\Lambda_n<0$ for $n=2,3,\dots,N$.
Then, the solution of the master equation is written as 
\be
 |p(t)\rangle = |R_1\rangle+\sum_{n=2}^N c_n e^{\Lambda_n t}|R_n\rangle, 
\ee
where $c_n$ is determined from the initial condition.
The bound in Eq.~(\ref{csl0}) is written as
\be
 && \frac{1}{2}\sum_{n=1}^N\int_0^t ds\,
 \left|\langle n|W|p(s)\rangle\right|  \no\\
 &=& \frac{1}{2}\sum_{n=1}^N\int_0^t ds\,
 \left|\sum_{m=2}^Nc_me^{\Lambda_ms}\Lambda_m\langle n|R_m\rangle\right| \no\\
 &\le& \frac{1}{2}\sum_{n=1}^N \sum_{m=2}^N\left(1-e^{\Lambda_mt}\right)
 \left|c_m\langle n|R_m\rangle\right|.
\ee
As we expect, the largest nonzero eigenvalue 
basically determines the relaxation processes.

This result shows that the bound in Eq.~(\ref{csl0}) 
converges to a finite value at the limit $t\to\infty$.
This property is contrasted to the case of quantum systems 
where the integrand in Eq.~(\ref{sl0}) or Eq.~(\ref{sl}) 
does not vanish at $t\to\infty$.
In that case, the bound exceeds 
the possible maximum value of the distance at the limit.

\subsection{Time-dependent stationary states}

When the transition-rate matrix $W(t)$ is a function of $t$, 
we can define the instantaneous stationary solution 
satisfying $W(t)|p^{\rm st}(t)\rangle=0$.
When the system is operated slowly,
the system follows the instantaneous stationary solution.
Then, the mechanism of the dynamics is different from 
that of the relaxation process.
We study implications from the speed-limit inequalities.

When the system is coupled to an external reservoir characterized by 
the inverse temperature $\beta$, 
the stationary state is given by the Gibbs distribution
\be
 p_n^{\rm st} = \frac{e^{-\beta E_n}}{\sum_{m=1}^Ne^{-\beta E_m}},
 \label{gibbs}
\ee
where $E_n$ denotes the energy of the state $|n\rangle$.
Both $\beta$ and $E_n$ can be time dependent.

We start the time evolution from the stationary state 
$|p(0)\rangle=|p^{\rm st}(0)\rangle$.
The time derivative of the Gibbs state $|p^{\rm st}(t)\rangle$ 
is formally written as 
\be
 |\dot{p}^{\rm st}(t)\rangle 
 = \left(W(t)+W_0^{\rm cd}(t)\right)|p^{\rm st}(t)\rangle, \label{cd0}
\ee
where 
\be
 W_0^{\rm cd}(t)
 =|\dot{R}_1(t)\rangle\langle L_1|
 =|\dot{p}^{\rm st}(t)\rangle\sum_{n=1}^N\langle n|.
\ee
We note that $|R_1(t)\rangle=|p^{\rm st}(t)\rangle$.
As we give a generalized discussion below, 
$W_0^{\rm cd}(t)$ is interpreted as a counterdiabatic term used in
the method of shortcuts to 
adiabaticity~\cite{DR03, DR05, Berry09, CRSCGM, STA13, STA19}.
It is introduced when we want to use the stationary state as 
the solution of the master equation.

In the present setting, we can apply the bound in Eq.~(\ref{csl0})
for $D(p(t),p(0))$.
By using the general formula in Eq.~(\ref{csl}), 
we can also examine the distance between the time-evolved state
and the instantaneous stationary state at each time.
We set $(p^{(1)}(t),p^{(2)}(t))=(p(t),p^{\rm st}(t))$ to obtain 
\be
 D(p(t),p^{\rm st}(t))
 &\le& \frac{1}{2}\sum_{n=1}^N\int_0^t ds\,
 \left|\dot{p}_n^{\rm st}(s)\right| \no\\
 &\le& \frac{1}{2}\int_0^t ds\,\sqrt{J^{\rm st}(s)}, \label{cslst}
\ee
where $J^{\rm st}(s)$ represents the Fisher information metric 
for $|p^{\rm st}(s)\rangle$.
When we drive the system slowly in time, 
the state $|p(t)\rangle$ is close to 
the instantaneous stationary state $|p^{\rm st}(t)\rangle$ and 
Eq.~(\ref{cslst}) is used to evaluate 
the adiabatic error, the deviation from the stationary state.

We note that the bound in Eq.~(\ref{cslst}) is represented by using
the stationary state only and is different from 
the bound in Eq.~(\ref{csl0}) for relaxation processes.
In Eq.~(\ref{int}), the relaxation dynamics is reflected
in the time-evolution operator $U^{(1)}$.
It is neglected in the course of the derivation of Eq.~(\ref{cslst}), 
which becomes an origin of a loose bound as we discuss below.
Although the stationary state is used as the bound, Eq.~(\ref{cslst})
is applicable to any state $|p(t)\rangle$.

The bound by the Fisher information metric of the stationary state 
is instructive since we can estimate the bound
from the nonequilibrium thermodynamic properties of the system.
When the system is driven slowly, the geometric thermodynamic length is  
represented by the Fisher information 
metric~\cite{Wootters81, SNB85, Crooks07}.
Suppose we discretize $t$ into $T$ time steps
$t_i=i\Delta t$ ($i=0,1,\dots,T$) with $t_T=t$. 
The time step $\Delta t$ is taken so that at each time $t_{i}$,
the system reaches the Gibbs state $|p^{\rm st}(t_{i})\rangle$,
Eq.~(\ref{gibbs}). 
The Kullback--Leibler divergence between the Gibbs distributions
at $t_{i}$ and $t_{i+1}$ is the total entropy production generated
during this time step as 
\be
 \Delta\sigma(t_i)
 &=& \sum_{n=1}^Np_n^{\rm st}(t_{i})
 \ln \frac{p_n^{\rm st}(t_{i})}{p_n^{\rm st}(t_{i+1})} \no\\
 &=& S(p^{\rm st}(t_{i+1}))-S(p^{\rm st}(t_{i}))-\beta\Delta Q(t_i),
\ee
where 
$S(p)=-\sum_{n=1}^Np_n\ln p_n$ is the Shannon entropy and 
$\Delta Q(t_i)=\sum_{n=1}^N(p_n^{\rm st}(t_{i+1})-p_n^{\rm st}(t_{i}))E_n(t_{i+1})$
is the heat absorbed from the environment. 
At the same time, by expanding up to the leading order in $\Delta t$, 
we obtain, $\Delta\sigma(t_i)\sim\Delta t^2J^{\rm st}(t_i)/2$, 
where $J^{\rm st}(s)$ represents
the Fisher information metric for $|p^{\rm st}(s)\rangle$. 
Therefore, the total entropy production is 
\be
 \Delta\sigma = \sum_{i=0}^{T-1}\Delta\sigma (t_i)
 \sim \frac{\Delta t}{2}\int_0^t ds\,J^{\rm st}(s).  
\ee
Practically, $\Delta t$ is approximately the relaxation time. 
Thus, we can estimate the bound of $D(p(t),p^{\rm st}(t))$
from the entropy production for quasistatic processes.

The bound of $D(p(t),p^{\rm st}(t))$ is basically applicable to 
any $|p(t)\rangle$ with 
the initial condition $|p(0)\rangle=|p^{\rm st}(0)\rangle$.
When the initial state is not equal to the instantaneous stationary state, 
it is convenient to use the quasistatic state instead of the stationary one.
The quasistatic state corresponds to the adiabatic state 
defined in closed quantum systems.
We write the transition-rate matrix as in Eq.~(\ref{spectral}).
In the present case, the eigenvalues and the eigenstates are time dependent.
We use the condition $\langle L_n(t)|\dot{R}_n(t)\rangle=0$ to fix the gauge.
Then, the quasistatic state is written as 
\be
 |p^{\rm qs}(t)\rangle=|R_1(t)\rangle+\sum_{n=2}^N 
 c_ne^{\int_0^t ds\,\Lambda_n(s)}|R_n(t)\rangle. \label{qs}
\ee
The stationary state $|p^{\rm st}(t)\rangle$ is obtained 
as a special case by setting $c_n=0$.
The time evolution of the quasistatic state is given 
by a similar form as Eq.~(\ref{cd0}): 
\be
 |\dot{p}^{\rm qs}(t)\rangle=\left(W(t)+W^{\rm cd}(t)\right)|p^{\rm qs}(t)\rangle.
\ee
The counterdiabatic term is given by
\be
 W^{\rm cd}(t)=\sum_{n=1}^N |\dot{R}_n(t)\rangle\langle L_n(t)|. \label{cd}
\ee
When we set $(p^{(1)}(t),p^{(2)}(t))=(p(t),p^{\rm qs}(t))$, we obtain 
\be
 D(p(t),p^{\rm qs}(t))
 &\le& \frac{1}{2}\sum_{n=1}^N\int_0^t ds\,\left|
 \langle n|W^{\rm cd}(s)|p^{\rm qs}(s)\rangle
 \right| \no\\
 &\le& \frac{1}{2}\int_0^t ds\,\sqrt{J^{\rm qs}(s)},
 \label{cslqs}
\ee
where $J^{\rm qs}(s)$ represents a generalization of 
the Fisher information metric
\be
 J^{\rm qs}(s)=\sum_{n=1}^N p_n^{\rm qs}(s)
 \left(\frac{\langle n|W^{\rm cd}(s)|p^{\rm qs}(s)\rangle}
 {p_n^{\rm qs}(s)}\right)^2.
\ee
This result is a classical counterpart of the quantum speed limit 
derived in Ref.~\cite{ST}.
It is known that the counterdiabatic term represents
a geometrical contribution of dynamics~\cite{Takahashi17}.

\section{Distance from the initial state in relaxation and annealing processes}
\label{sec:app1}

We discussed the bound of $D(p(t),p(0))$ as in Eq.~(\ref{csl0}).
It is also bounded from above by thermodynamic quantities~\cite{SFS}.
When we have the detailed balance condition
\be
 \langle m|W|n\rangle=\e^{-\beta(E_m-E_n)}\langle n|W|m\rangle,
 \label{dbc}
\ee
we can show 
\be
 D(p(t),p(0))\le \frac{1}{2}\int_0^t ds\,\sqrt{2\dot{\sigma}(s) A(s)},
 \label{cslthermo}
\ee
where $\dot{\sigma}$ is the entropy production rate 
\be
 \dot{\sigma}=\frac{1}{2}\sum_{m\ne n}\left(
 W_{mn}p_n-W_{nm}p_m\right)\ln\frac{W_{mn}p_n}{W_{nm}p_m} \label{ep}
\ee
and $A$ is the activity
\be
 A=\frac{1}{2}\sum_{m\ne n}\left(
 W_{mn}p_n+W_{nm}p_m\right).
\ee

It is instructive to study which of  
Eqs.~(\ref{csl0}) and (\ref{cslthermo}) gives a tight bound.
We study the simplest two-state system described by the general 
form of the transition-rate matrix 
\be
 W(t)=\frac{k(t)}{2}\bmat{cc} -(1-r(t)) & 1+r(t) \\
 1-r(t) & -(1+r(t)) \emat, \label{w2}
\ee
where $k(t)$ is a nonnegative function and 
$r(t)$ satisfies $-1\le r(t)\le 1$.
The stationary state is given by 
\be
 |p^{\rm st}(t)\rangle =\frac{1}{2}\bmat{c} 1+r(t) \\ 1-r(t) \emat.
\ee

\begin{figure}[t]
\begin{center}
\includegraphics[width=1.0\columnwidth]{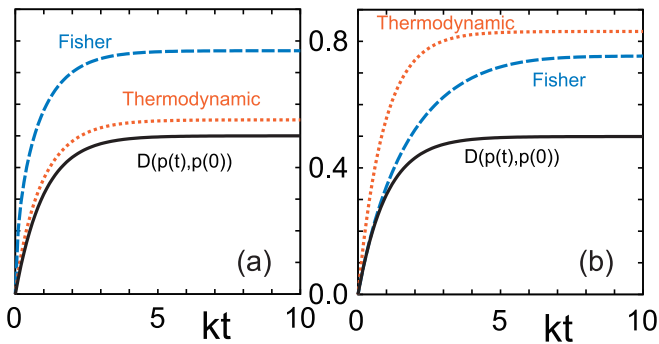}
\caption{Relaxation processes for two-state systems.
We set the time independent parameters 
for the transition-rate matrix in Eq.~(\ref{w2}).
In each panel, we plot $D(p(t),p(0))$ (black solid line), 
the bound by the Fisher information metric 
in the rightmost side of Eq.~(\ref{csl0}) (red dashed line),  
and the thermodynamic bound in the right hand side of Eq.~(\ref{cslthermo})
(blue dotted line).
We set the initial state $(p_1(0),p_2(0))=(1,0)$ and 
the stationary state $(p_1^{\rm st},p_2^{\rm st})=(0.5,0.5)$
in panel (a), and $(p_1(0),p_2(0))=(0.5,0.5)$ and 
$(p_1^{\rm st},p_2^{\rm st})=(0.999,0.001)$ in panel (b).
}
\label{fig1}
\end{center}
\end{figure}

When we consider a time-independent $W$, the system shows a relaxation
behavior to the stationary state.
The eigenvalues of Eq.~(\ref{w2}) are given by 0 and $-k$,
and the nonzero eigenvalue, $-k$, determines the relaxation scale.
We show numerical results in Fig.~\ref{fig1}.
As we see from the figure, 
the bound in Eq.~(\ref{csl0}) can be better 
than that in Eq.~(\ref{cslthermo}), and vice versa.

The equality condition of the Cauchy--Schwartz inequality 
used in the last line of Eq.~(\ref{csl0}) 
is given by $|\dot{p}_n(t)|\propto p_n(t)$.
In panel (a) of Fig.~\ref{fig1}, we start the time evolution from 
the distribution with $p_1\gg p_2$
toward the uniform distribution with $p_1\sim p_2$.
Since the relation $|\dot{p}_1|=|\dot{p}_2|$ holds in the two-state system,
the equality condition can be satisfied only at large $t$.
Then, the bound gives a loose one at small $t$.
On the other hand, the initial distribution is given 
by the uniform one in panel (b), 
and we observe a small deviation of the bound from the distance
for small $t$.

The thermodynamic bound in Eq.~(\ref{cslthermo}) depends on 
the entropy production rate in Eq.~(\ref{ep}).
It becomes large when 
the difference $|W_{mn}p_n-W_{nm}p_m|$ for $m\ne n$ takes a large value.
In panel (b) of Fig.~\ref{fig1}, $W_{21}/k=(1-r)/2$ takes a small value.
As a result, the thermodynamic bound gives a loose one.

We next study annealing processes.
We set the instantaneous stationary state as Eq.~(\ref{gibbs}) 
and start the time evolution from $|p(0)\rangle=|p^{\rm st}(0)\rangle$.
We fix the energy levels and change the inverse temperature $\beta(t)$
as function of time.
We increase the inverse temperature linearly in $t$ from 
$\beta(0)=0$ to $\beta(t_{\rm f})=\beta_{\rm f}$.

\begin{figure}[t]
\begin{center}
\includegraphics[width=1.0\columnwidth]{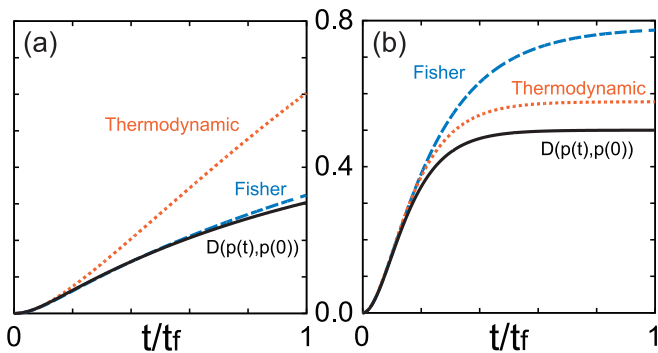}
\caption{Annealing processes for two-state systems.
The instantaneous stationary state is given by  
$p_n^{\rm st}(t)=e^{-\beta(t) E_n}/\sum_me^{-\beta(t) E_m}$
and the initial condition is $|p(0)\rangle=|p^{\rm st}(0)\rangle$.
We set the energy levels 
$\beta_{\rm f} (E_1,E_2)=(0,20)$
and the protocol $\beta(t)/\beta_{\rm f}=t/t_{\rm f}$.
See the caption of Fig.~\ref{fig1} for plotted lines.
(a) A nonadiabatic process with $\beta_{\rm f}k=1$.
(b) An adiabatic process with $\beta_{\rm f}k=10$.
}
\label{fig2}
\end{center}
\end{figure}

The results are basically dependent on the scale $k$ in 
the transition-rate matrix.
When the relaxation scale $k$ is much smaller than
the annealing scale $\beta_{\rm f}(E_2-E_1)/t_{\rm f}$, the state  
is changed before the relaxation and $|p(t)\rangle$ significantly 
deviates from the instantaneous stationary state $|p^{\rm st}(t)\rangle$.
In the opposite limit, 
the system quickly relaxes to the instantaneous stationary state 
and we find $|p(t)\rangle\sim |p^{\rm st}(t)\rangle$.
As we see numerical results in Fig.~\ref{fig2}, 
the bound by the Fisher information metric gives a tight bound
at the former regime and 
that by the thermodynamic quantities gives a tight bound at the latter.

Thus, we conclude that the bounds 
in Eqs.~(\ref{csl0}) and (\ref{cslthermo}) can be useful 
as different measures.
The differences of their bounds from the actual distance  
depend strongly on the condition of the process.
We note that the thermodynamic bound in Eq.~(\ref{cslthermo}) is applicable
only when we use the detailed balance condition.
The bound by the Fisher information metric is applicable to any processes.
It can be used when we study the distance between 
$|p(t)\rangle$ and $|p^{\rm st}(t)\rangle$
as we discuss in the next section.
It is also known that the Fisher information metric is 
related to the entropy production rate by an inequality~\cite{ID20}.


\section{Distance from the stationary state in annealing processes}
\label{sec:app2}

The speed-limit inequalities for $D(p(t),p(0))$
in Eqs.~(\ref{csl0}) and (\ref{cslthermo})
are not practically useful
since the bound is represented 
by using the time-evolved state $|p(t)\rangle$.
When we study $D(p(t),p^{\rm st}(t))$, the bound is represented 
by $|p^{\rm st}(t)\rangle$ only and becomes a convenient relation 
when we want to know the behavior of the unknown state $|p(t)\rangle$ 
from the known state $|p^{\rm st}(t)\rangle$.
In this section, we treat annealing processes and examine 
the relation in Eq.~(\ref{cslst}).
We note that the relation using the thermodynamic quantities as 
Eq.~(\ref{cslthermo}) is not known in this case.

The system takes $N$ states and the set of energy levels 
$\{E_n\}_{n=1,2,\dots, N}$ satisfies the relation
$E_1< E_2\le E_3\le \cdots \le E_N$.
We change the temperature of the reservoir from 
$\beta(0)\sim 0$
to $\beta(t_{\rm f})\to\infty$ monotonically as a function of $t$.
The initial state is given by the uniform distribution
$p_n(0)=1/N$ and 
we expect $p_n(t_{\rm f}) \sim \delta_{n,1}$ at 
the processing time $t=t_{\rm f}$.
When we operate the system very slowly, 
the state at each $t$ is close 
to the Gibbs distribution in Eq.~(\ref{gibbs}).

\begin{figure}[t]
\begin{center}
\includegraphics[width=1.0\columnwidth]{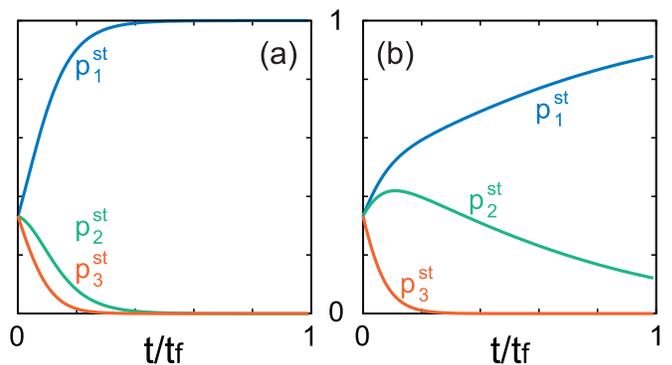}
\caption{The instantaneous stationary distribution of Eq.~(\ref{gibbs})
for three-state systems.
We use the linear protocol $\beta(t)/\beta_{\rm f}=t/t_{\rm f}$
with $0\le t\le t_{\rm f}$.
The behavior of each probability distribution is determined by 
the energy level.
(a) We set $\beta_{\rm f}(E_1,E_2,E_3)=(0,12,20)$.
Each function is given by a monotonic function.
(b) We set $\beta_{\rm f}(E_1,E_2,E_3)=(0,2,20)$.
$p_2^{\rm st}(t)$ changes its behavior at
$t=t_2^*$ where $t_2^*$ is determined from $E_2=\bar{E}(t_2^*)$.
}
\label{fig3}
\end{center}
\end{figure}

The bound of $D(p(t),p^{\rm st}(t))$ is given by Eq.~(\ref{cslst}).
The right hand side of the first line is dependent 
on the monotonicity of the Gibbs distribution.
The ground state 
$p_1^{\rm st}(t)$ is a monotonically increasing function
of $t$
and the highest state 
$p_N^{\rm st}(t)$ is a monotonically decreasing function.
The other states are dependent on each energy level.
$p_n^{\rm st}(t)$ is an increasing function 
when $E_n$ is larger than the average value 
$\bar{E}(t)=\sum_{m=1}^N E_mp_m^{\rm st}(t)$, 
and decreasing when $E_n<\bar{E}(t)$.
Since $\bar{E}(t)$ is a monotonically decreasing function, 
we have several possible patterns as shown in Fig.~\ref{fig3}.
The right hand side of the first line in Eq.~(\ref{cslst}) is written as 
\be
 \frac{1}{2}\sum_{n=1}^N\int_0^t ds\,
 \left|\dot{p}_n^{\rm st}(s)\right| 
 = \sum_{n=1}^N\max_{t_n^* (0\le t_n^*\le t)}
 p_n^{\rm st}(t_n^*) -1. 
 \label{sl1}
\ee
In the simplest case $E_1\le \bar{E}(t)\le E_2\le E_3\le E_N$, 
the bound can be estimated by using the ground state as 
\be
 \frac{1}{2}\sum_{n=1}^N\int_0^t ds\,
 \left|\dot{p}_n^{\rm st}(s)\right| 
 = p_1^{\rm st}(t)-p_1^{\rm st}(0). 
\ee
In the other cases, we have additional positive contributions,
which gives a loose bound.
This property is reasonable since the annealing procedure 
works well when the ground state energy is small enough compared to 
the other ones.
The case in panel (b) is interpreted as a hard problem 
compared to that in panel (a).
We note that this criterion is not related to 
the structure of local minima 
presented by the quasifree energy~\cite{KGV83}.
The structure is reflected to 
the detailed form of the transition-rate matrix.
The hardness of the problem represented by the speed limit is
determined by the energy level structure.

We examine the bound in the second line of Eq.~(\ref{cslst}).
It is written as 
\be
 \frac{1}{2}\int_0^t ds\,\sqrt{J^{\rm st}(s)}
 &=& \frac{1}{2}\int_0^t ds\,|\dot{\beta}(s)|
 \Delta E(\beta(s)) \no\\
 &=& \frac{1}{2}\int_{\beta(0)}^{\beta(t)} d\beta\,\Delta E(\beta), \label{sl2}
\ee
where $\Delta E$ represents the energy fluctuation 
\be
 \Delta E(\beta(s))=\sqrt{\sum_{n}E_n^2p_n^{\rm st}(s)
 -\left(\sum_{n}E_np_n^{\rm st}(s)\right)^2}.
\ee
The second line of Eq.~(\ref{sl2}) is obtained when $\beta(s)$ 
is a monotonously increasing function of $s$.
It shows that the bound is independent of the processing speed
and is determined only by the initial and final values, which is consistent
with the fact that the bound is a geometrical quantity and is independent 
on the details of the annealing procedure.

In the case of three-state systems,
the transition-rate matrix satisfying the stationary condition
$W(t)|p^{\rm st}(t)\rangle =0$
with the detailed balance condition is generally parametrized as 
\be
 W=\bmat{ccc}
 -(\pi_2 a+\pi_3 b) & \pi_1 a & \pi_1b \\
 \pi_2 a & -(\pi_1 a+\pi_3 c) & \pi_2 c \\
 \pi_3 b & \pi_3 c & -(\pi_1 b+\pi_2c)
 \emat.
 \no\\
 \label{w3}
\ee
where we put $\pi_n=p_n^{\rm st}(t)$, 
and the symbols $a$, $b$, and $c$ represent nonnegative parameters.

\begin{figure}[t]
\begin{center}
\includegraphics[width=1.0\columnwidth]{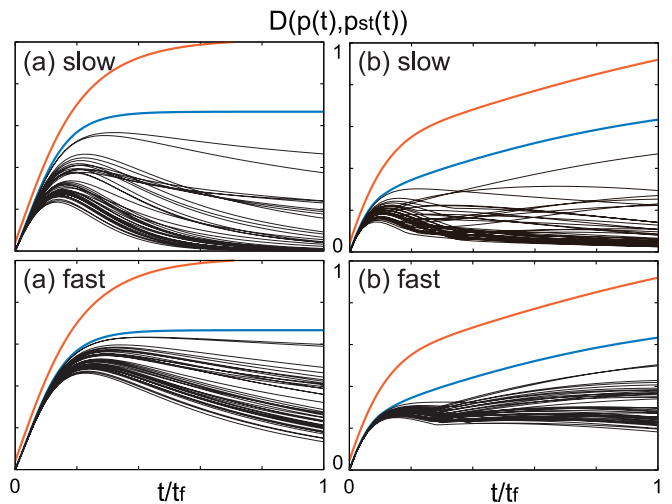}
\caption{
Annealing processes for three-state systems.
We adopt the energy levels and the protocols used in Fig.~\ref{fig3}.
We use the protocol in Fig.~\ref{fig3}(a) for the left panels,
and the protocol in Fig.~\ref{fig3}(b) for the right panels.
We generate random numbers for parameters $a$, $b$, and $c$
to plot 40 samples in each panel. 
We set $at_{\rm f}, bt_{\rm f}, ct_{\rm f}\le 10$ in the upper panels
with the caption ``slow'',
and $at_{\rm f}, bt_{\rm f}, ct_{\rm f}\le 2$ in the lower panels
with ``fast''.
In each panel, 
the lower thick solid line (blue) represents Eq.~(\ref{sl1})
and the upper thick solid line (red) represents Eq.~(\ref{sl2}).
The thin solid lines (black) are plotted by solving the master equation 
numerically with the transition-rate matrix in Eq.~(\ref{w3}).
}
\label{fig4}
\end{center}
\end{figure}

We randomly generate the transition-rate matrix and 
calculate the distance $D(p(t),p^{\rm st}(t))$.
The result is plotted in Fig.~\ref{fig4}.
We see that Eq.~(\ref{sl1}) gives a tight bound for a small $t$ 
while we find a large deviation for a large $t$.
We also see that the fast driving with a small $t_{\rm f}$
gives a tight bound compared to the slow driving with a large $t_{\rm f}$.
We note that the bounds are independent on the choice of $t_{\rm f}$.
The bound discussed in the speed limit
describes the worst case evaluation and 
the distance becomes much smaller than the bound at a large $t$.
This is because the effect of the relaxation dynamics described 
by the nonzero eigenvalues of $W$ is completely neglected 
in Eq.~(\ref{cslst}).
Furthermore, 
each component of the state $|p(t)\rangle$ is restricted to 
a nonnegative value.
This is contrasted to the quantum case where 
each component of the state takes a complex value and 
shows an oscillating behavior as a function of $t$.

\section{Pumping processes}
\label{sec:app3}

\subsection{General results for two-state systems}

As a possible application of the derived bounds, 
we finally treat pumping processes.
The system is coupled to multiple reservoirs and 
is driven periodically.
When the driving protocol satisfies a geometrical condition,
especially at the slow-driving regime, 
we have a nontrivial current, 
which is known as the Thouless pumping~\cite{Thouless, SN}.
In this system, the state settles down to a periodic behavior 
after transient evolutions at the first several periods.
Since the speed-limit inequality generally gives a loose bound
at large processing times, we need to devise an efficient method 
to treat periodically-driven systems.

The property of the pumping processes can be understood 
by using the two-state system.
Before discussing the pumping processes, 
we study the general properties of the two-state system~\cite{TFHH}.
The transition-rate matrix is generally written as Eq.~(\ref{w2}).
The solution of the master equation is written as
\be
 |p(t)\rangle &=& \frac{1}{2}\bmat{c} 1+r(t) \\ 1-r(t) \emat \no\\
 && +\left(c\e^{-\int_0^t ds\,k(s)}+\delta(t)\right)\bmat{c} 1 \\ -1 \emat,
 \label{ptwo}
\ee
where $c$ is determined from the initial condition and 
$\delta(t)$ is obtained by solving the differential equation 
\be
 \dot{\delta}(t)=-k(t)\delta(t)-\frac{1}{2}\dot{r}(t)
 \label{ddelta}
\ee
with the initial condition $\delta(0)=0$.
We note that $\delta(t)$ satisfies $|\delta(t)|\le 1$.
The quasistatic state in Eq.~(\ref{qs}) 
is obtained by setting $\delta(t)=0$ for Eq.~(\ref{ptwo}), 
which shows the relation 
\be
 D(p(t),p^{\rm qs}(t))=\left|\delta(t)\right|.
\ee

In the two-state system, the counterdiabatic term 
defined generally by Eq.~(\ref{cd})
is calculated to give
\be
 W^{\rm cd}(t)=\frac{1}{2}\dot{r}(t)\bmat{cc} 1 & 1 \\ -1 & -1 \emat.
\ee
Then, we obtain the bound in the first line of Eq.~(\ref{cslqs}) as 
\be
 D(p(t),p^{\rm qs}(t)) \le \frac{1}{2}\int_0^t ds\ |\dot{r}(s)|. \label{b2}
\ee
This inequality is understood from the formal integral representation
of $\delta(t)$: 
\be
 \delta(t)=-\frac{1}{2}\int_0^t ds\,e^{-\int_s^t du\,k(u)}\dot{r}(s).
 \label{delta}
\ee
This exact form involves 
an exponentially-decaying factor, which is neglected in Eq.(\ref{b2}).
As a result the bound is expected to give a tight bound when
the parameter $k(t)$ takes a small value.
In the pumping processes, the situation is realized 
when we operate the system rapidly.

When the system oscillates rapidly, it is expected that 
the improved bound given generally in Eq.~(\ref{csl2}) gives a
better result.
In the two-state case, it is given by
\be
 D(p(t),p^{\rm qs}(t)) &\le& 
 \frac{1}{2}\left|r(t)-r(0)\right| \no\\
 && +\frac{1}{2}\int_0^t ds\,k(s)\left|r(s)-r(0)\right|.
\ee

\subsection{Pump current}

When the system is coupled to the left and the right reservoirs,
the transition-rate matrix is decomposed as 
\be
 W(t)=\sum_{\alpha=L,R}
 \bmat{cc} -k_{\rm in}^{\rm (\alpha)}(t) & k_{\rm out}^{\rm (\alpha)}(t) \\
 k_{\rm in}^{\rm (\alpha)}(t) & -k_{\rm out}^{\rm (\alpha)}(t) \emat.
\ee
$k_{\rm in}^{(\alpha)}$ and $k_{\rm out}^{(\alpha)}$ 
denote the incoming amplitude and the outgoing amplitude 
from the system to the reservoir $\alpha$ respectively,
when the first/second component of $|p(t)\rangle$ represents
the probability that the state is empty/filled.
Then, the current from the system to the right reservoir is defined as 
\be
 J = \lim_{T\to\infty}\frac{1}{T}\int_{0}^{T} dt\,
 \left(k_{\rm out}^{\rm (R)}(t)p_2(t)
 -k_{\rm in}^{\rm (R)}(t)p_1(t)\right). 
\ee
Substituting Eq.~(\ref{ptwo}) to this expression, we find that 
the current consists of the dynamical and 
the geometrical contributions.
The latter is given by 
\be
 J_{\rm g}=\lim_{T\to\infty}\frac{1}{T}\int_{0}^{T} dt\,
 k^{\rm (R)}(t)\delta(t),
\ee
where $k^{\rm (R)}(t)=k_{\rm out}^{\rm (R)}(t)+k_{\rm in}^{\rm (R)}(t)$.
It is represented by a flux 
penetrating a surface in a parameter space~\cite{SN, TFHH}.

When the system is driven periodically with the period $T_0$, 
$\delta(t)$ is written as 
\be
 \delta(t)=\tilde{\delta}(t)-\tilde{\delta}(0)e^{-\int_0^t ds\,k(s)},
 \label{tdeltadef}
\ee
where the first term represents a periodic function satisfying 
$\tilde{\delta}(t+T_0)=\tilde{\delta}(t)$ 
and the second term describes a transient decaying behavior.
$\tilde{\delta}(t)$ obeys the same differential equation as $\delta(t)$
and the initial value $\tilde{\delta}(0)$ is obtained from
\be
 \delta(T_0)=\tilde{\delta}(0)\left(1-e^{-\int_0^{T_0} dt\,k(t)}\right).
 \label{delta0}
\ee
We can write the geometric current as 
\be
 J_{\rm g}=-\frac{1}{T_0}\int_{0}^{T_0} dt\,k^{\rm (R)}(t)\tilde{\delta}(t).
\ee
Combining Eqs.~(\ref{delta}), (\ref{tdeltadef}), and (\ref{delta0}),
we obtain 
\be
 && \tilde{\delta}(t)
 = -\frac{1}{2}\int_0^t ds\, e^{-\int_s^t du\,k(u)}\dot{r}(s) \no\\
 && -\frac{1}{2}\frac{e^{-\int_0^{T_0} ds\,k(s)}}{1-e^{-\int_0^{T_0} ds\,k(s)}}
 \int_0^{T_0} ds\, e^{-\int_s^t du\,k(u)}\dot{r}(s). \label{tdelta}
\ee
This is upperbounded as
\be
 \left|\tilde{\delta}(t)\right|
 \le \frac{1}{2}\left(1+\frac{1}{\int_0^{T_0} ds\,k(s)}\right)
 \int_0^{T_0} ds\,\left|\dot{r}(s)\right|, \label{bound1}
\ee
and 
\be
 && \left|\tilde{\delta}(t)\right|
 \le \frac{1}{2}\left|r(t)-r(0)\right| \no\\
 &&
 +\frac{1}{2}\left(1+\frac{1}{\int_0^{T_0} ds\,k(s)}\right)
 \int_0^{T_0} ds\,\left|r(s)-r(0)\right|. \label{bound2}
\ee
The latter is obtained
by using the integration by parts of Eq.~(\ref{tdelta}).
By using these bounds, we can evaluate the geometric current as 
\be
 |J_{\rm g}|\le\frac{1}{T_0}\int_{0}^{T_0} dt\,k^{\rm (R)}(t)
 \left|\tilde{\delta}(t)\right|. \label{bj}
\ee

\begin{figure}[t]
\begin{center}
\includegraphics[width=0.80\columnwidth]{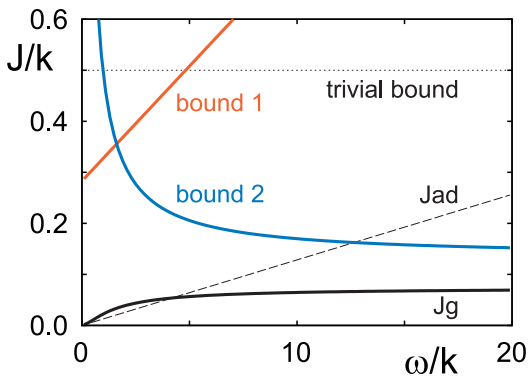}
\caption{Geometric current $J_{\rm g}$ and its bounds as functions
of the frequency $\omega=2\pi/T_0$.
We set $k(t)=k$ as a constant, 
$r(t)=0.2\times[\sin (\omega t)+\cos (\omega t)]$, and 
$k^{\rm (R)}(t)/k=0.5\times(1+0.8\times\sin(\omega t-0.6\pi))$.
The red line (bound 1) denotes the bound (\ref{bj}) with Eq.~(\ref{bound1})
and the blue line (bound 2) denotes that with Eq.~(\ref{bound2}).
We also show the trivial bound 
$|J_{\rm g}|\le\frac{1}{T_0}\int_{0}^{T_0} dt\,k^{\rm (R)}(t)$
by the dotted line.
The dashed line $J_{\rm ad}$ denotes the geometric current obtained 
by using the adiabatic approximation.
}
\label{fig5}
\end{center}
\end{figure}

We show a numerical result in Fig.~\ref{fig5}.
The bound in Eq.~(\ref{bound1}) is basically an increasing function 
with respect to the frequency $\omega=2\pi/T_0$.
As a result the bound becomes loose when we increase $\omega$.
On the other hand, it gives a finite contribution at $\omega\to 0$.
Since the geometric current is known to vanish at the limit,
bound 1 cannot be a good approximation to the geometric current.
Concerning the second bound in Eq.~(\ref{bound2}), 
we observe that the bound is a decreasing function of $\omega$.
Although it gives a loose bound for small $\omega$, 
the asymptotic behavior for large $\omega$ gives a better result.
Since the function $r(t)$ is an oscillating function, 
it is not possible for the bound to give a saturated result.
We see that the bound is several times larger than the actual value 
of the current.

\section{Summary and discussions}
\label{sec:summary}

We have discussed speed-limit inequalities for 
classical stochastic processes described by the master equation.
For a distance between arbitrary time-evolved states, 
the bound is constructed from a geometrical perspective.
We find that the bound is represented by 
using the Fisher information metric and its generalization.
Our main general results are represented
in Eqs.~(\ref{csl}) and (\ref{cslf}).
By using these results, we can obtain
Eqs.~(\ref{csl0}), (\ref{cslst}), and (\ref{cslqs}).

The distance $D(p(t),p(0))$ is a useful measure 
when we discuss relaxation dynamics.
For thermodynamic systems with detailed balance condition, 
the distance is also bounded from above by
the nonequilibrium thermodynamical quantities.
We compared the bound by the Fisher information metric 
and that by the thermodynamic quantities in simple examples 
and found that no general inequality holds between two bounds. 
The former can be larger or smaller than the latter.
It strongly depends on the condition of the process.

We can also study the bound of the distance $D(p(t),p^{\rm st}(t))$ 
by using the Fisher information metric.
It is useful when we study annealing dynamics 
of the instantaneous stationary state.
When we change the stationary state as a function of time,
the state follows the instantaneous stationary state.
We show that the speed limit inequality gives
the worst case evaluation of the annealing processes.

Since the speed-limit inequality is universally applied to any systems,
the result often gives a loose bound.
As we see from the derivation of the inequality in Eq.~(\ref{u}),
it is impossible to realize the saturating condition in the inequality.
The second term of the right hand side in Eq.~(\ref{int})
has components with different signs, which is obstacle to the tight bound.
This is contrasted to the quantum speed limit inequality where
we can find nontrivial time evolutions keeping the saturating condition.
In the present classical stochastic systems,
the state $|p\rangle$ represents a probability and
each component cannot be negative,
which gives the difference from the quantum case.

In spite of this problem, the derived speed limit for classical stochastic
systems is applied to a broad class of processes.
From a practical perspective, it is important to find a useful bound 
which can be evaluated without knowing the details of the system.
Since our result is applicable to the distance 
$D(p^{(1)}(t),p^{(2)}(t))$ between arbitrary states,
it is in principle possible to find 
a more useful relation by choosing the states 
in a proper way~\cite{Takahashi22}.

\section*{Acknowledgments}
We are grateful to Satoshi Nakajima for useful discussions.
This work was supported by 
JSPS KAKENHI Grants No. JP20H01827, No. JP20K03781, and No. JP20H05666.


\end{document}